\documentclass[preprint, 12pt]{elsarticle}

\usepackage{lineno,hyperref}
\modulolinenumbers[5]
\usepackage{setspace}
\usepackage{color}
\usepackage{subfigure}
\usepackage{amsmath}
\journal{Carbon}

\begin{document}

\begin{frontmatter}

\title{Carbon nanotubes collapse phase diagram with arbitrary number of walls. Collapse modes and macroscopic analog.}

\author[myaddress1,myaddress2]{Y. Magnin\corref{mycorrespondingauthor1}}
\cortext[mycorrespondingauthor1]{Corresponding author}
\ead{magnin@mit.edu}
\author[myaddress3]{F. Rondepierre}
\author[myaddress4]{W. Cui}
\author[myaddress5]{D.J. Dunstan}
\author[myaddress3]{A. San-Miguel\corref{mycorrespondingauthor2}}
\cortext[mycorrespondingauthor2]{Corresponding author}
\ead{alfonso.san-miguel@univ-lyon1.fr}

\address[myaddress1]{MIT Energy Initiative, Massachusetts Institute of Technology, Cambridge, MA, United States}
\address[myaddress2]{Consultant, Total@Saclay NanoInnov, 2 boulevard Thomas Gobert, 91120 Palaiseau Cedex, France.}
\address[myaddress3]{Univ Lyon, Universit\'e Claude Bernard Lyon 1, CNRS, Institut Lumi\`ere Mati\`ere, Campus LyonTech - La Doua, F-69622 LYON, France}
\address[myaddress4]{School of Physics and Electronic Engineering, Jiangsu Normal University, Xuzhou 221116,China}
\address[myaddress5]{School of Physics and Astronomy, Queen Mary University of London, London, E1 4NS, UK}
\date{\today}

\doublespacing

\begin{abstract}
Carbon nanotubes tend to collapse when their diameters exceed a certain threshold, or when a sufficiently large external pressure is applied on their walls. Their radial stability of tubes has been studied in each of these cases, however a general theory able to predict collapse is still lacking. Here, we propose a simple model predicting stability limits as a function of the tube diameter, the number of walls and the pressure. The model is supported by atomistic simulations, experiments, and is used to plot collapse phase diagrams. We have identified the most stable carbon nanotube, which can support a maximum pressure of $\sim$18 GPa before collapsing. The latter was identified as a multiwall tube with an internal tube diameter of $\sim$12nm and $\sim$30 walls. This maximum pressure is lowered depending on the internal tube diameter and the number of walls. We then identify a tube diameter domain in which the radial mechanical stability can be treated as equivalent to macroscopic tubes, known to be described by the canonical L\'evy-Carrier law. This multiscale behavior is shown to be in good agreement with experiments based on O-ring gaskets collapse, proposed as a simple macroscopic parallel to nanotubes in this domain.
\end{abstract}

\end{frontmatter}

\begin{keyword}
Carbon nanotubes, High pressure, Irreversible transformation, Atomistic simulations
\end{keyword}


\section{Introduction}
Low-dimensional carbon structures such as fullerenes, graphene, carbon nanotubes (CNT), nanocones, nano-junctions~\cite{kroto,geim,iijima,yang,wei,nasir} have deeply changed fundamental concepts of condensed matter physics during the last decades~\cite{dinadayalane}. The many associated technological breakthroughs have opened perspectives in a broad range of applications, ranging from electronics~\cite{peng2014,shulaker2013}, sensor developments~\cite{chen2016},energy transport and storage~\cite{halakoo,kim,saito} or biology and medical sciences such as drug delivery technology~\cite{bianco}. While both graphene and CNT sp$^2$ structures have concentrated an important part of the recent research efforts, there exist many hybrid structures between these two, which have been much less explored ~\cite{he2019}. In the literature, they are referred as "collapsed nanotubes", "flattened carbon nanotubes", "closed-edge graphene nanoribbons" or "dogbones". Such structures correspond to a geometrical evolution of the CNT radial cross-section, from circular towards a continuum of shapes, in which the internal walls become closer, in at least one radial direction. The terms mentioned above are most frequently used when the distance between the internal walls in the collapse direction tends to the graphitic interlayer distance, where van der Waals interactions (vdW) have to be considered. For clarity, we will refer to this state as the "collapsed" shape, and the deviations from the circular cross-section leading to it as "collapse transition" shapes. The latter can be either first-order-like for large tube diameters, or continuous for smaller ones~\cite{2005-Tangney}, going through different geometries including oval, race-track or polygonal~\cite{2001-Rols}, all grouped in the "collapse transition" domain.\\
Characterizing the CNT collapse behavior is greatly motivated by the change in electronic structure from the pristine circular cross-section to the deformed or collapsed geometry~\cite{2000-Lammert,2000-Mazzoni,2000-Yang, 2002-Gulseren,2020-Impellizzeri,giusca,balima}. Hence, a geometrical electronic tuning based on a shape modification may offer an interesting alternative to substitutional doping in nano-engineering design~\cite{2002-Duclaux,ma,2018-Machon-Perspective}.
\begin{figure*}[htp]
    \begin{center}
	\includegraphics[width=1.0\linewidth]{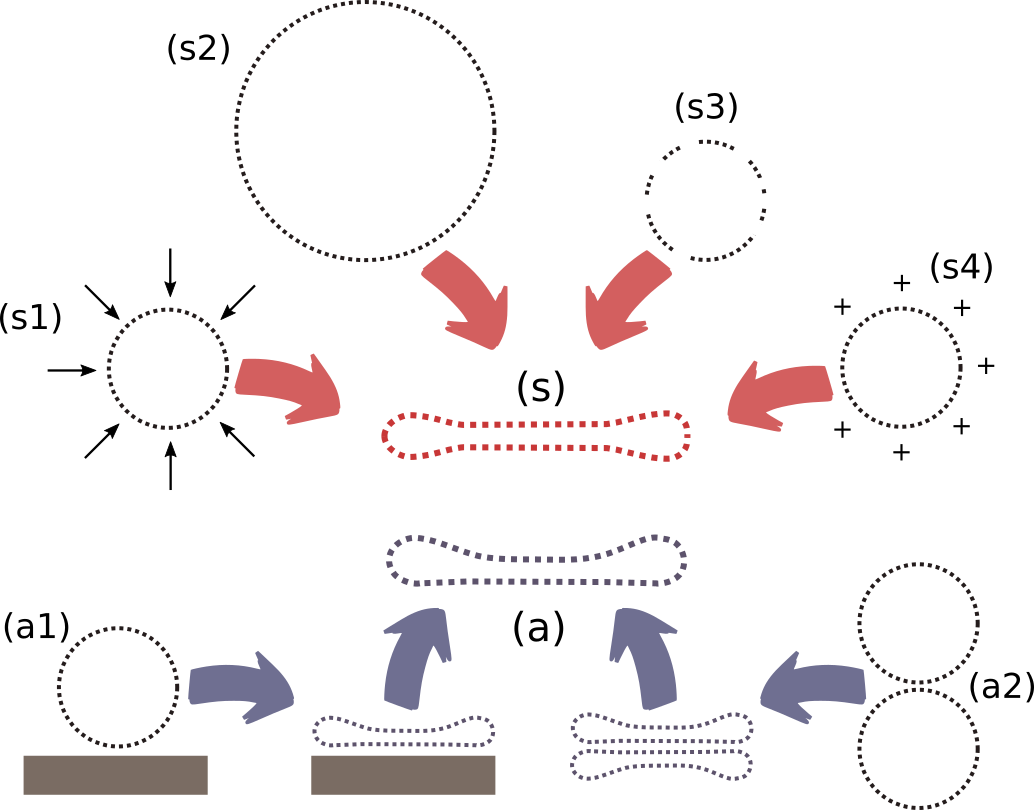}
	\caption{Scheme of the different physical mechanisms allowing the evolution of a carbon nanotube to a symmetrical collapsed structure (s): \textbf{s1} External applied pressure; \textbf{s2} Self collapse for large tube diameters; \textbf{s3} Defective tubes; \textbf{s4} Through charge injection. Mechanisms leading to an asymmetrical collapsed structure (a): \textbf{a1} Through interaction with a substrate; \textbf{a2} Through interaction with other nanotubes.}
	\label{fig0}
    \end{center}
\end{figure*}
Soon after the first dedicated study about CNT by Iijima et al.~\cite{1991-Iijima}, collapsed geometries were evidenced by electron microscopy on large CNT diameters~\cite{1995-Chopra}. It is generally admitted now that a collapse event is favored for large tube diameters and/or for small numbers of concentric tubes~\cite{1998-Benedict,2005-Tang,2010-Pugno,2012-Zhang-defective-collapse,2016-Balima}. Other collapse parameters at ambient conditions were also identified, including interactions between the external walls of CNT, interactions with a substrate or with molecular adsorbates~\cite{1998-Hertel,2009-Yan,2010-Xie}, defect formation by electron beam irradiation ~\cite{1996-Chopra}, or due to the application of electrostatic fields~\cite{2011-Shklyaev,2016-Barzegar}. The different mechanisms that can be responsible for collapse are illustrated in Figure \ref{fig0}.\\
For small tube diameters with a stable circular cross-section at ambient conditions, it has been shown that a high external pressure causes collapse. This was first shown for SWCNT, where the collapse pressure has been demonstrated to be a function of CNT diameter~\cite{1999-Sood,2003-Chan,2004-Capaz,2004-Elliott,2005-Merlen,2005-Tangney,2006-Zhang,2006-Hasegawa,2008-Caillier,2008-Yao,2009-Ghandour,2010-Kuntscher,2013-Sun,2014-Cerqueira,2017-Torres-Dias}. Pressure-induced collapse is also observed for double-wall~\cite{2005-Ye,2006-Gadagkar,2011-Aguiar,2011-You}, triple-wall~\cite{2017-Alencar} and multi-wall carbon nanotubes~\cite{2009-Shima}. Any effect of helicity (or chirality) - a geometrical characteristic reported to originate from the configurational tube edge entropy~\cite{magnin} - on the pressure response of CNT was reported to be small, or to occur only in tubes of very small diameters~\cite{2015-Torres-Dias}.\\
In this work, we explore the stability conditions of CNT from their circular cross-section to collapsed shapes, focusing on the stability domain as a function of geometrical parameters (diameter and number of tube walls), as well as the effect of external pressure. The instability of the circular cross-section is found to be driven by the competition of the elastic energy, related to the bond-bending energy, and the external and internal forces. External forces include the pressure applied to CNT external surfaces, and interactions with the surrounding molecular medium, while internal forces include the vdW interactions of the tube walls. While a number of atomistic calculations and models have tried to predict the CNT stability conditions~\cite{2004-Capaz,2004-Sluiter,2005-Tangney,2005-Ye,2006-Gadagkar,2006-Hasegawa,2009-Shima,2010-Pugno,2014-Cerqueira,2017-Torres-Dias,2004-Sluiter}, none of them fully covered structural tube properties, including tube diameters, number of walls, etc.\\
In addition, we propose a multiscale approach, showing that the collapse onset of a range of nanotubes with diameter about nanometer sizes, are well described by the canonical L\'evy-Carrier law (LC), formulated 150 years ago~\cite{carrier}, and originally developed for macroscopic tube collapse. Our modified LC-based model behaves consistently in this domain  with atomistic simulations, density functional tight-binding (DFTB), molecular dynamics (MD), as well as with a  macroscale analog based on O-ring gaskets deformation. Our model is then used to plot collapse phase diagrams for carbon nanotubes, providing a better understanding of the radial stability and collapse mechanisms of single- and multi-wall nanotubes (MWCNT), including nanotube bundles as a function of their diameter from the nanometer to dozen of nanometers. We finally think that such an approach could be adapted for more complex porous materials.

\section{Results and discussion}
\subsection{Theoretical SWCNT collapse model}
In mechanics, the radial collapse pressure $P_c$ for macroscopic tubes with a diameter $d_0$, is known to scale as $P_c\ \propto d^{-3}_{0}$, as expressed by L\'evy-Carrier~\cite{carrier}. At the nanoscale, it has been shown that such a formalism is consistent for SWCNT, when including an additional correction term, $\beta^2/d^2_{0}$. This term emerges both in simulations, as well in experiments, and may be related to the large built-in curvature energy for small tube diameters, or to the discrete nature of the nanotubes~\cite{carter2020softening}. This approach is called the modified LC equation~\cite{2017Torres-Dias}, and is written,
\begin{equation}
\label{eq:eq1}
P_c = \frac{24 D}{d^3_{0}} \left(1- \frac{\beta^2}{d^2_{0}} \right),
\end{equation}
where $D$ is the bending stiffness of graphene, and $\beta$ corresponds to the diameter of the smallest free-standing stable SWCNT~\cite{2017Torres-Dias}. All parameters are given in Table \ref{tbl:LC-parameters}. When $d_0 \gg \beta$, the LC law is recovered; however, when $d_0 < \beta$, $P_c<$0, corresponding to unfeasibly small unsupported tube diameters.\\
Eq.\ref{eq:eq1}, originally based on experimental observations~\cite{2017-Torres-Dias}, has been shown to be well-suited for SWCNT with diameters in the range $d_0 \sim$~0.7~-~2~nm, and was found to be independent of the tube chirality~\cite{2017Torres-Dias}. With increasingly larger tube diameters, the correction term decreases, and the interaction of the external tube with the pressure transmitting medium (PTM) becomes more important. To include the PTM in the model, we integrate Eq.\ref{eq:eq1}, and add a surface energy $\gamma_{F-C}$, in order to account for the surrounding PTM (an argon bath in MD simulations)~\cite{2012-Pugno}. Doing so, we obtain the enthalpy of a SWCNT of length $L$ as,
\begin{equation}
\label{eq:eq2}
H\ =\ \frac{48 D}{d_0^3}\left( 1- \frac{\beta^2}{3d_0^2} \right)\ \frac{\pi L}{4} d^2_0\ +\ P\frac{\pi L}{4} d^2_0+\ \gamma_{F-C}\ \pi L\ d_0,
\end{equation}
Minimizing Eq.\ref{eq:eq2} as a function of $d_0$, we obtain the collapse pressure for SWCNT interacting with the PTM as,
\begin{equation}
\label{eq:eq3}
P_c\ = \ \frac{24 D}{d^3_0}\left(1- \frac{\beta^2}{d^2_{0}} \right)\ -\ 2\frac{\gamma_{F-C}}{d_0}.
\end{equation}

\subsection{Theoretical SWCNT bundle collapse model}
For a bundle formed of SWCNT, Eq.\ref{eq:eq3} can be modified following Pugno et al.~\cite{2012-Pugno}, considering that each individual tube in a bundle interacts with its neighboring tubes, acting as a SWCNT pseudo-fluid, while the outer surface of the bundle interacts with the PTM. Then, $P_c$ can be obtained by minimizing the corresponding enthalpy expression,
\begin{equation}
\label{eq:eq4}
P_c\ = \ \frac{24 D}{d^3_0}\left(1- \frac{\beta^2}{d^2_{0}} \right)\ -\ 2 \left( \frac{\gamma_{C-C}}{d_0}\ +\ \frac{\gamma_{F-C}}{d_B}\right)
\end{equation}
where d$_B$ represents the bundle diameter. In Eq.\ref{eq:eq4}, $\gamma_{C-C}$ is the carbon formation energy, corresponding to the inter-tube vdW interactions, while the last term represents the interaction between the PTM and the external surface area of the bundle. It is worth to note that following Pugno et al.~\cite{2010-Pugno} and consistent with atomistic modeling~\cite{2017-Alencar}, the bundle will have undergone polygonization before collapse.
\subsection{Theoretical MWCNT collapse model}
To go a step further, we now generalize our model for MWCNT. We follow the methodology proposed by Gadakar et al.~\cite{2006-Gadagkar} in which friction between tubes is neglected so that the bending stiffnesses are additive. Thus, the net external pressure needed to collapse the $N$ walls of a MWCNT is written as the sum of the pressures $P_{c_i}$ needed to collapse the $i=1$ to $N$ corresponding individual SWCNT. The pressure energy needs to be distributed between the different tubes, leading to the additive character of the $P_{c}$. In MWCNT, the vdW inter-wall interactions are compensated, considering that each individual inner-tube interacts both with its next inner- and next outer-tube, $i$-1 and $i$+1 respectively. However, it is necessary to account for interaction of the innermost tube ($i$=0) with only its next larger tube neighbor ($i$=1), and for the interaction of the outermost tube ($i$=$N$-1) with ($i$=$N$-2), and finally with the external PTM (the last term in Eq.\ref{eq:eq5}). Accounting for all these interactions, the collapse pressure becomes,
\begin{equation}
\label{eq:eq5}
\begin{split}
P_c &  = \ \sum^{N-1}_{i=0}\ \frac{24 D}{d_i^3}\ \left( 1- \frac{\beta^2}{d_i^2} \right)\ -\ 2\left(\ \frac{\gamma_{C-C}}{d_{0}}\ -\ \frac{\gamma_{C-C}}{d_{N-1}}\ +\ \frac{\gamma_{F-C}}{d_{N-1}}\right).
\end{split}
\end{equation}
Inter-tube distances are reported to range in between 0.27~nm and 0.42~nm, however, the most common distances in MWCNT are about 0.32-0.35~nm~\cite{Khari2014}. For the sake of simplicity, we have considered that all tubes in a MWCNT range at the graphitic interlayer distance $\delta$ (see Table.\ref{tbl:LC-parameters}), and that $d_i= d_0+ 2\delta \cdot i$ in Eq.\ref{eq:eq5}. We may note here that even if a general MWCNT can no longer be considered a thin tube, our method of evaluation of the collapse pressure as contribution of various SWCNT in interaction with their environment allows us to keep using the LC expression, which is in fact only valid for thin-walled tubes.\\

\begin{table}
  \caption{Parameters used in the various L\'evy-Carrier models.}
  \label{tbl:LC-parameters}
  \begin{center}
  \begin{tabular}{lll}
  	\hline
  	\hline
    $D$             & \hspace{1cm} 1.7 (eV)        & \hspace{1cm} Ref~\cite{2017Torres-Dias,meng}\\
    \hline
    $\beta$         & \hspace{1cm} 0.44 (nm)        & \hspace{1cm} Ref~\cite{2017Torres-Dias}\\
    \hline
    $\gamma_{F-C}$  & \hspace{1cm} 0.11 (J/m$^2$)  & \hspace{1cm} This work\\
    \hline
    $\gamma_{C-C}$  & \hspace{1cm} 0.23 (J/m$^2$)  & \hspace{1cm} Ref~\cite{meng,han,girifalco}\\
    \hline
    $\delta$        & \hspace{1cm} 0.34 (nm)        & \hspace{1cm} Ref~\cite{meng}\\
    \hline
  	\hline
  \end{tabular}
  \end{center}
\end{table}

\subsection{Numerical simulations, experiments and model validation}
\begin{figure*}[htp]
    \begin{center}
	\includegraphics[width=1.0\linewidth]{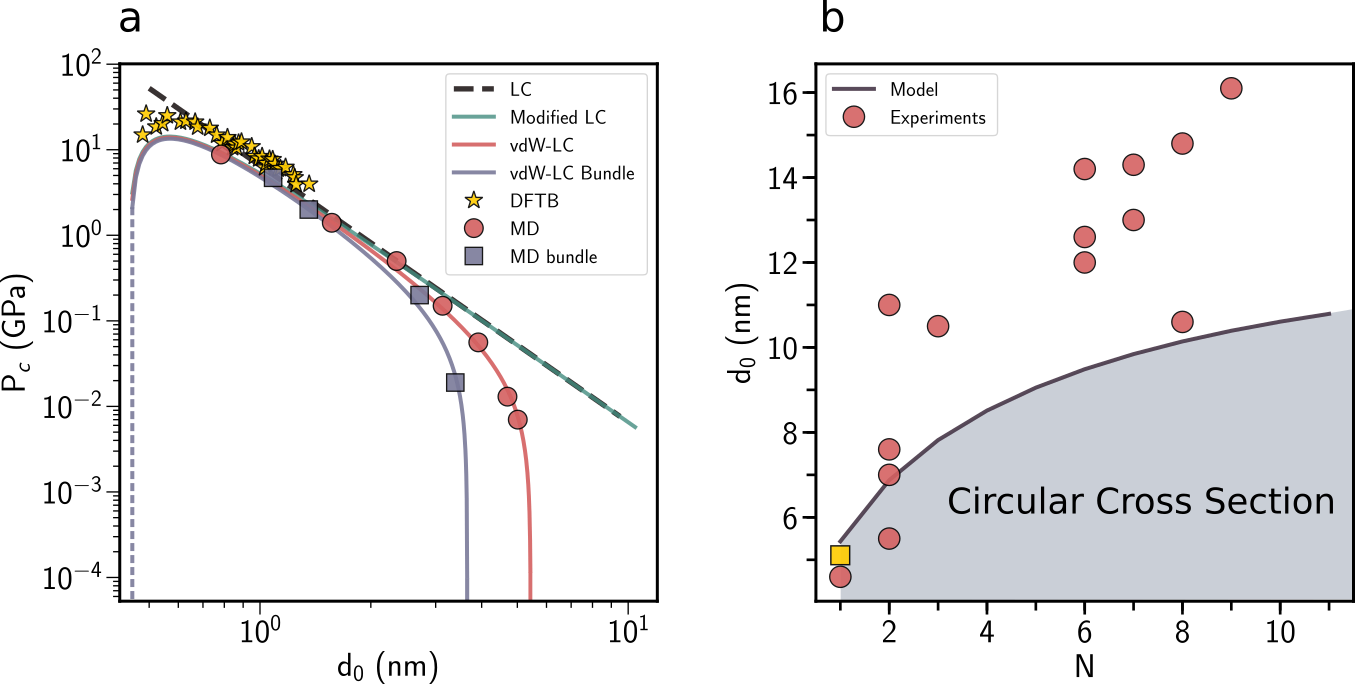}
	\caption{\textbf{a.} Collapse pressure $P_c$ of SWCNT as a function of the tube diameter $d_0$, using three models based on the L\'evy-Carrier formula: the standard LC (black dashed line), the modified LC (green line) and the vdW-LC developed in this work for SWCNT (red line), and bundles (blue line). The results are compared to numerical simulations, DFTB (yellow stars) and MD simulations with a long-range bond order potentials for SWCNT (red circles), and bundles (blue squares). \textbf{b.} Innermost tube diameters $d_0$ of MWCNT for which collapsed configurations have been observed, plotted against the number of tube walls $N$. The red circles correspond to experimental observation of collapsed tubes, while the yellow square corresponds to an experimental determination of the collapse pressure for SWCNT. The black line corresponds to the prediction from our vdW-LC model. The gray area corresponds to the phase where tubes are not collapsed (circular cross section), while the part of the diagram above corresponds to nanotubes collapsed at ambient pressure.}
	\label{fig1}
    \end{center}
\end{figure*}

In order to check the accuracy of our model, we compare it with both numerical and experimental data. Simulations have been performed with two algorithms, the DFTB for small tube diameters~\cite{aradi2007}, and MD based on the empirical AIREBO bond order potential~\cite{stuart}, accounting both for C-C covalent bonds and for long-range vdW interactions~\cite{magnin2}. All simulations have been performed for tubes or bundles immersed in an Ar bath at 300K. Ar-Ar and C-Ar interactions have been modeled by a (12-6) Lennard-Jones potential, using the Lorentz-Berthelot mixing rule (see section Method for simulation details).\\
In Fig.\ref{fig1}.a, we show the evolution of $P_c$ as a function of the tube diameter from DFTB, MD, and from the theoretical models detailed above. As can be seen, for small $d_0$, the modified LC approach (light green line), is in good agreement with the DFTB simulations (yellow stars). Hence, when $d_0<$0.57~nm, $P_c$ decreases, corresponding to a situation where tube diameters are so small that the curvature energy is large enough to make tubes unstable. It is noteworthy that a deviation is observed when it is compared to the macroscopic LC law (dashed black line), that does not include the small-diameter correction term discussed above. When $d_0>$3nm, the vdW-LC approach (red line) shows a sharp decrease of $P_c$, corresponding to a tube diameter where interactions between the PTM and the tube walls dominate, while the curvature energy is negligible for such diameters. SWCNT simulations (red circles), show a self-collapse diameter of 5.3~nm, in good agreement with the experimental data reported; about 5.1~nm is cited in the review of He et al.~\cite{he2019}. This agreement allows then to fit $\gamma_{F-C}$ (see Table \ref{tbl:LC-parameters}), in the vdW-LC model for SWCNT. As shown in Fig.\ref{fig1}.a, we found an excellent agreement between simulations and the theoretical model for the full diameter range simulated.
We then compared the vdW-LC model applied on bundles. In Fig.\ref{fig1}.a, the gray squares correspond to the results of simulations performed on bundles made of 37 SWCNT. As can be seen, the self-collapse diameter is smaller than that of isolated SWCNT, a behavior already observed in~\cite{2007-Motta}. This behavior could be explained by the contribution of the  polygonisation of the tubes to the surface energy which results from inter-tube interactions and which tends to lower $P_c$. Using Eq.\ref{eq:eq4} and the $\gamma_{F-C}$ previously fitted from isolated SWCNT, we see that the vdW-LC model applied to the bundle configuration is in good agreement with simulations using the C-C interaction energy $\gamma_{C-C}$ (see Table \ref{tbl:LC-parameters}). Note that in the bundle used in simulations, $d_B\gg d_0$ and the interaction between the PTM and the external bundle surface could be neglected compared to the inter-tube interactions.\\
In Fig.\ref{fig1}.b, we compare experiment with the vdW-LC for MWCNT (Eq.\ref{eq:eq5}). The tube stability is presented as innermost tube diameter $d_0$ against the number of tube walls $N$ at ambient pressure. The continuous line separating the two domains represents the critical internal diameter for collapse of MWCNT at ambient pressure. The experimental points (red circles) correspond to collapsed tubes observed by electron microscopy, and are extracted from different sources summarized in the work of Balima et al~\cite{2016-Balima}. As expected, these points are mainly found in the collapsed domain. We have underlined a particular point from the work of He et al~\cite{he2019} (yellow square), which corresponds to the determination of the critical collapse diameter by  observations of many SWCNT. This point should then lie on the limiting curve, and is found to be in very good agreement with the prediction of our model. We can also note that two points are found bellow the theoretical prediction. These exceptions may be explained by interactions with the substrate (Fig.\ref{fig0}.a1), which are known to favor the tube collapse~\cite{2013-Blancon}. Overall, our prediction is very consistent with experiments.\\
The result presented in  Fig.\ref{fig1}.b also shows that with an increasingly higher number of tube walls, MWCNT of larger internal diameter can be stabilized at ambient conditions. Note that the inner-wall vdW interactions may lead to the metastability of a collapsed geometry at a diameter below the critical one~\cite{1995-Chopra,2007-Motta,2010-Pugno,2016-Balima}. The critical diameter may also be strongly reduced for defective carbon nanotubes~\cite{2012-Zhang-defective-collapse}.

\subsection{Nanotube collapse phase diagram}
\begin{figure*}[htp]
    \begin{center}
	\includegraphics[width=1.0\linewidth]{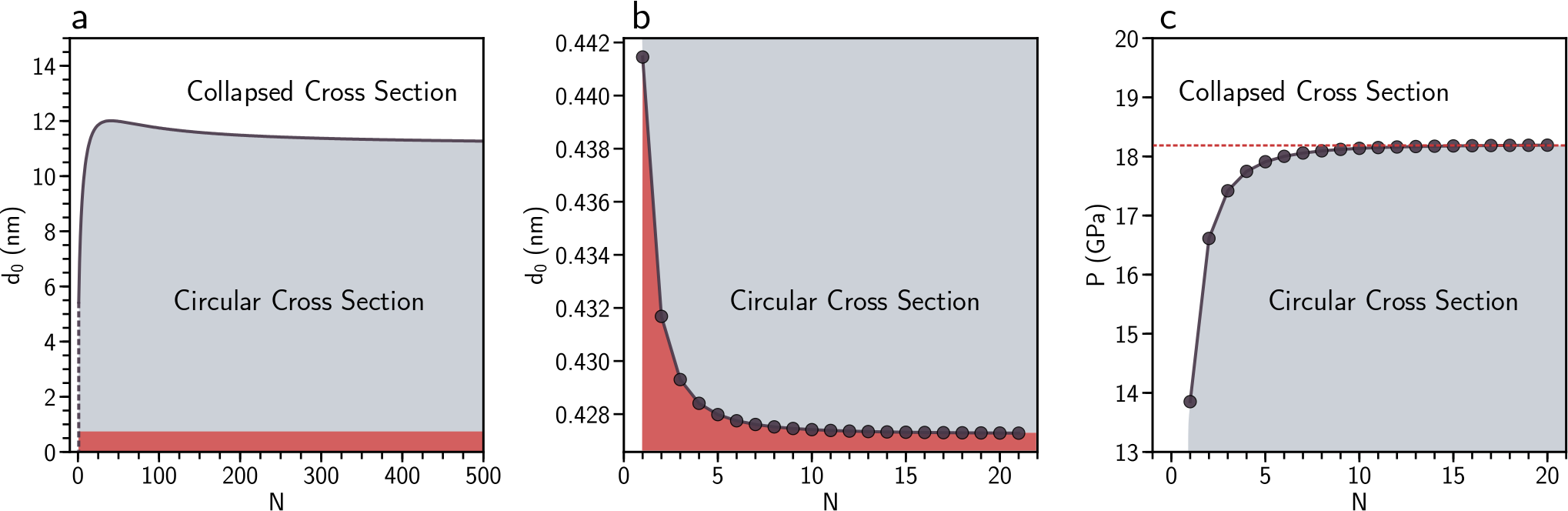}
	\caption{\textbf{a}. Stability diagram of carbon nanotubes at ambient pressure as a function of the internal diameter $d_0$, and the number of tube walls, $N$. Three regions are defined from the bottom to the top: (1) the red phase corresponds to the tubes that cannot exist due to too small an internal diameter, (2) the gray phase, corresponding to the stability domain of tubes with a circular cross-section, and (3) the white phase, corresponding to the stability domain of collapsed tubes. \textbf{b}. Zoom on the red phase in (a) (unstable tubes), for small tube diameters. \textbf{c}. Stability diagram for MWCNT with $d_0$=0.56~nm, corresponding to the tubes found to show the highest collapse pressure. The red dashed line corresponds to the maximum collapse pressure found in the model.}
	\label{fig2}
    \end{center}
\end{figure*}

We have demonstrated that the vdW-LC model is a robust, simple and suitable approach in predicting single- and multi-wall carbon nanotubes stability, as well as for bundles. We now use this model to determine the nanotube collapse phase diagram. In Fig.\ref{fig2}.a, we plot the stability diagram of MWCNT at ambient pressure, extending beyond the domain explored in Fig.~\ref{fig1}.b. The stable phase, either circular or collapsed, depends on $d_0$ and $N$. There also exists an instability domain, i.e., in which tubes cannot exist, for very small $d_0$ (red domain) in Fig,\ref{fig2}.a,b. Hence, with an increasingly large $d_0$, the number of tube walls has to be increased in order to stabilize a circular MWCNT. Nevertheless, $d_0$ has a maximum at $d_0$=12nm for $N$=45 walls, corresponding theoretically to the largest possible internal cavity in MWCNT. For larger $N$, $d_0$ slightly decreases and becomes asymptotically constant at $d_0$=11.3~nm. This behavior results from the competition between the inter-tube vdW and the tube-PTM interactions, corresponding to the second term in the Eq.\ref{eq:eq5}.\\

\noindent In  Fig.\ref{fig2}.b, we see that the smallest possible tube diameter for a free-standing SWCNT corresponding to $d_0$=0.44~nm~\cite{2017Torres-Dias} can be slightly reduced when increasing the number of tube walls. This result is qualitatively supported in the literature, where $d_0\sim$0.407nm has been reported for double-wall carbon nanotubes~\cite{Guan2008}, and $d_0\sim$0.3nm for a $N$=13 MWCNT~\cite{zhao2004}. Our calculations do not allow for internal tube diameters below $\sim$0.43nm. Such a limitation could results from the tube-substrate interactions, the discreet distribution of carbon atoms onto the wall, or the small tube diameter helicities, neglected in the model, and going beyond the scope of this work.\\
\begin{figure*}[htp]
    \begin{center}
	\includegraphics[width=1.\linewidth]{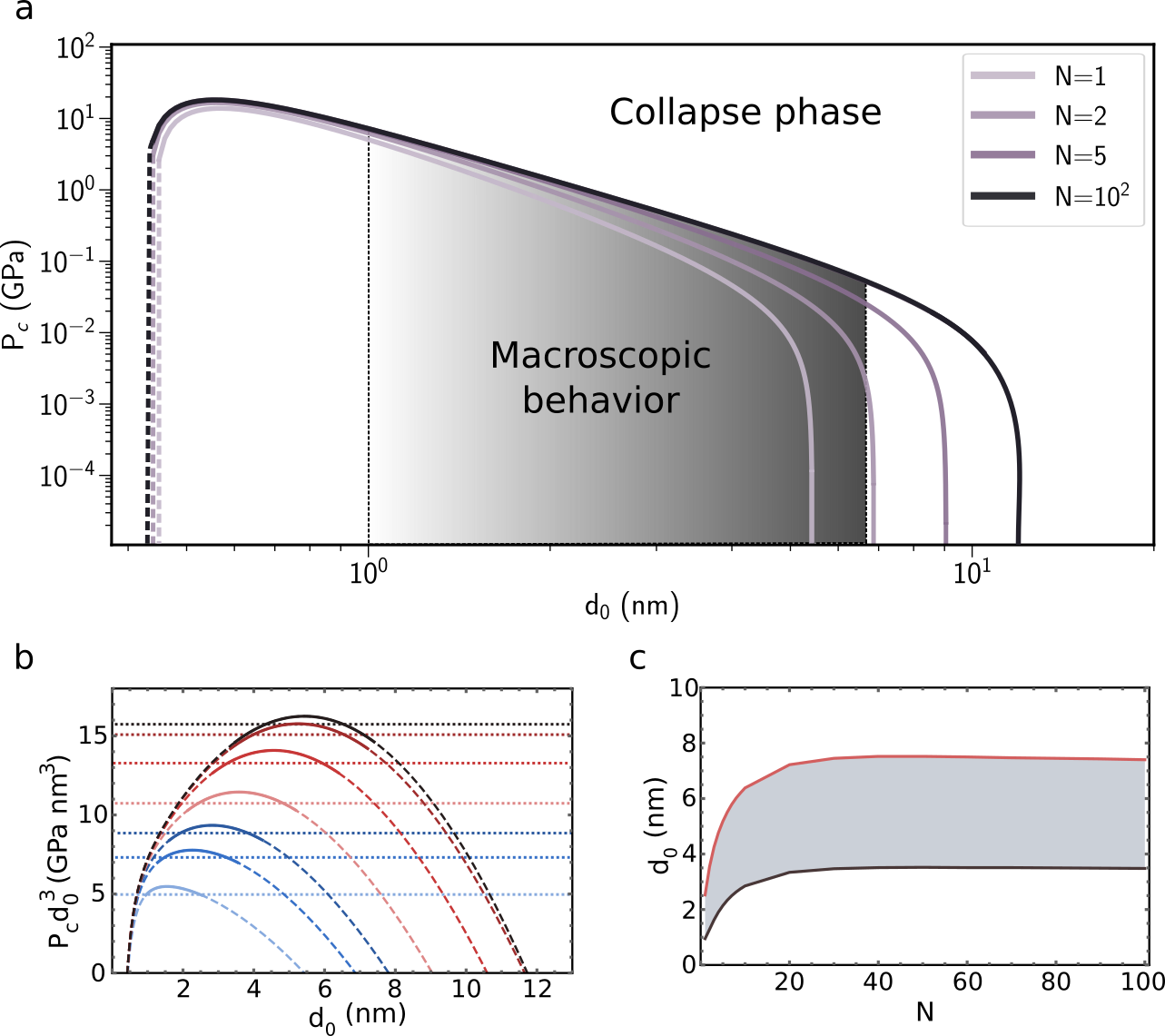}
	\caption{\textbf{a.} Collapse pressure $P_c$ as a function of the innermost tube diameter $d_0$ for different numbers of tube walls $N$, ranging from 1 to 100 and determined from the theoretical model accounting for vdW interactions. The gray area represents the multiscale domain, i.e. a $d_0$ range where the original macroscopic LC model agrees with the vdW-LC model at the nanoscale (see text and (b) for details and criteria). \textbf{b.} Normalized collapse pressure $P_c d_0^3$ as function of the internal diameter for tubes with $N$=1, 2, 3, 5, 10, 20 and 100 walls from inner to outer curves. The continuous part of the curves represent the domain in which within $\pm$5 \% accuracy, a macroscopic LC model (horizontal dashed lines) can be used to approximate the vdW-LC one.\textbf{c.} Domain of validity of the continuum mechanics corresponding to the LC model (gray area). We have considered that the LC model is valid when it differs by less than 5\% from a linear approximation in the $P_c d_0^3$ representation as function of $d_0$. }
	\label{fig3}
    \end{center}
\end{figure*}
\noindent The maximum pressure needed to collapse any MWCNT can now be found by searching the maximum collapse pressure from Eq.\ref{eq:eq5}, as a function of $d_0$ and $N$. The maximum  $P_c$ value depends on $N$ but is found invariably at $d_0\sim$0.56~nm whatever the value of $N$. The maximum value of $P_c$ evolves from $P_c\sim$13.9~GPa for SWCNT and progressively increases with $N$, converging to a maximum at $P_c$=18.2~GPa. As shown in Fig.\ref{fig2}.c, we note the quick convergence of $P_c$ for $N$ ranging from 1 to 4. The number of walls is thus found to increase the stability pressure by about $\sim$30\%, corresponding to the observations on deformed or collapsed geometries for MWCNT with a relatively high number of walls in  nanocomposites under compression~\cite{2016-Balima}.\\
We now use the model to find how $P_c$ evolves as a function of $d_0$, for $N$ ranging from 1 to 100, in the collapse phase diagram, Fig.\ref{fig3}.a. As can be seen, for small $d_0$, $N$ has no significant effect on $P_c$. So, we may approximate that all curves collapse in a single one for internal diameters below $d_0\sim$1nm. For larger $d_0$, $N$ plays a more significant role on tubes stability. Interestingly, in Fig.\ref{fig3}.b, the collapse pressures obtained with the LC model are comparable with those from the vdW-LC model for certain tubes. We consider a multiscale domain when the $P_c d_0^3$ plotted as function of $d_0$ can be approximated by the LC law corresponding to a constant. We show this behavior for $N=$ 1, 2, 3, 5, 10, 20 and 100, where the continuous part of plots can be assimilated to horizontal lines, assuming an uncertainty on experimental pressure determinations of the order of $\pm 5$ \%. In Fig~\ref{fig3}.c, we show the tube diameter as a function of the number of walls into the multiscale domain. We thus observe that SWCNT with diameters ranging from 0.95 to 2.5 nm behave as macroscopic tubes. This diameter domain is shifted as a function of $N$ to higher diameters, and converging for $N=$ 20, for internal diameters ranging from $\sim$3.5 to $\sim$7.5 nm. The chosen accuracy of $\pm 5$ \% corresponds to a conservative choice of pressure determination in high-pressure measurements~\cite{2008-Syassen}. This link between the nano- and the macroscopic scales can be understood from the dominant role played by the innermost tube on pressure stability, as shown in Fig.~\ref{fig2}.a. It is noteworthy that such a multiscale behavior has been previously reported in the case of SWCNT~\cite{2004-Capaz,2004-Zang}.

\section{A macroscopic model for the collapse of carbon nanotubes}
We have shown above that Eq.\ref{eq:eq1} is a suitable multiscale equation for comparing the collapse behavior of macroscopic tubes with certain tubes at the nanoscale. This result is thus consistent in predicting $P_c$ for SWCNT with diameters ranging from $d_0 \sim$ 1-2~nm and for MWCNT with a limited number of tube walls. Note that the diameter range reported corresponds to the most common tube diameters produced experimentally~\cite{Prasek2011}.
\begin{figure*}[htp]
    \begin{center}
	\includegraphics[width=1.0\linewidth]{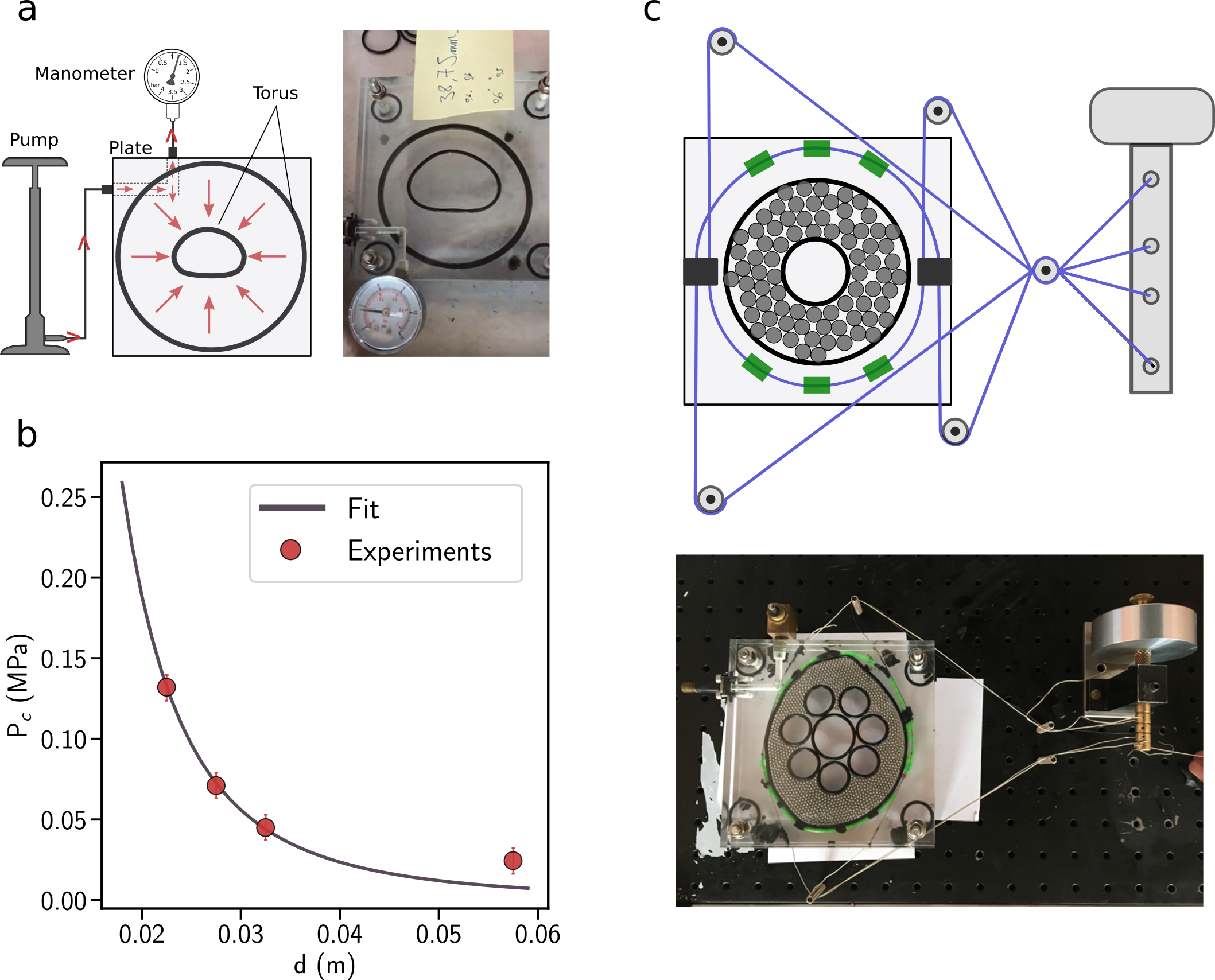}
	\caption{\textbf{a.} Schema and picture of the experiment, in which O-rings are collapsed by a fluid pressure medium (air or water). \textbf{b.} Averaged collapse pressure $P_c$ of O-rings in the experiment described in (a), as a function of their diameters $d$ (red circles). Experimental data are found to fit $P_c\propto d^{-3.0 \pm 0.7}$ (black line). \textbf{c.} Schema and picture of the experiment in which O-rings are collapsed by a solid  pressure medium (ball-bearings ).}
	\label{fig4}
    \end{center}
\end{figure*}
\noindent In order to verify this theoretical result, we have developed a table-top experiment using macroscopic nitrile rubber gaskets to mimic CNT. In this experiment, nanotubes are replaced by macroscopic toroidal rings (O-rings) with the diameters $d_0$=11.25, 13.7, 16.25 and 28.75~mm. The O-rings are placed between two transparent plates to limit movement in the direction of the torus axis. A pressure vessel is made by using a large outer O-ring surrounding the smaller inner one. The vessel is connected to a bicycle pump with a pressure gauge to generate and monitor the pressure acting on the inner O-ring (Fig.\ref{fig4}.a). The deformation of this O-ring is measured as a function of the applied pressure, and quantifieded by image analysis (details are given in the section Method. Videos of the experiment are also available in the Supplementary Information). The O-rings used in this experiment are not strictly speaking tubes, but rings, and we first verify that they follow the LC law. To do so, four measurements per O-ring diameters have been performed, and the averaged data are shown in Fig.\ref{fig4}.b. The data are fitted by an inverse power law, $P_c=a.D^{-\alpha}$ with $a$=1.51 J (a constant), and $\alpha$=3.0 $\pm$ 0.7, as expected from Eq.\ref{eq:eq1}. The O-rings are thus demonstrated to have a similar radial mechanical response to macroscopic tubes, and can reasonably be used for comparison with SWCNT with $d_0$ ranging from $\sim$1 to 3.45~nm.\\
In Fig.\ref{fig5}, we compare the dynamical collapse behavior of a macroscopic O-ring bundle with data from MD simulations at the nanoscale. In the experiment, a bundle of 37 O-rings is surrounded by a bath of steel ball-bearings acting as a PTM. Pressure is applied by tightening a sheathed guitar string surrounding the ball-bearings (Fig.\ref{fig4}.c). 
\begin{figure*}[htp]
	\centering
	\includegraphics[width=.9\linewidth]{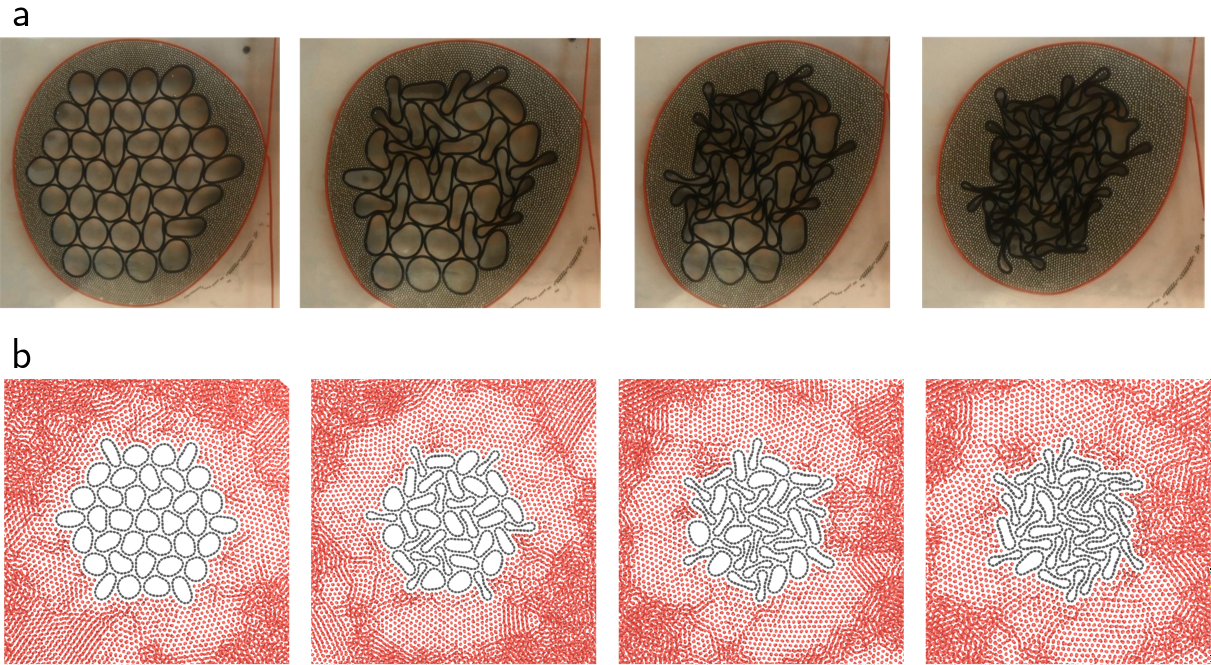}
	\caption{\textbf{a.} Evolution of an O-ring bundle immersed in the ball-bearing pressure-transmitting  medium during a compression cycle. \textbf{b.} Collapse behavior of a SWCNT bundle immersed in a solid argon medium. Simulation was performed at pressures ranging from 1 to 3 GPa with a pressure increased (from the left to the right).}
	\label{fig5}
\end{figure*}
The change of PTM (ball-bearings instead of air or water) compared to the previous experiment (Fig.\ref{fig4}.a) is necessary to match simulations. Indeed, the pressure needed to collapse a bundle with SWCNT diameter around 1~nm is a few GPa, a pressure at which the argon PTM used in simulation and in experiment is solid ~\cite{bolmatov}. The ball-bearings tend to form an hexagonal close-packed planar domain, which mimic a macro-crystalline argon PTM. In the simulation, we have built a bundle formed of 37 SWCNT with $d_0$=1.3nm, immersed in an argon bath at $P\sim$2GPa. Collapse experiments on O-rings are compared to these MD simulations in Fig.\ref{fig5}. The collapse process in both cases is found to be in good qualitative agreement. In the early stages, tubes show a small deformation. Then some of the tubes collapse to a peanut shape, while others stay circular or ovalize. Later in the semi-collapsed state, a large number of tubes have collapsed, but a few tubes remain ovalized. At the end of the process, all tubes show a collapse shape. Note that both experiments and simulations show that in a solid PTM, the collapse of a SWCNT bundle, at least when it can be assimilated to a macroscopic bundle, is a complex and non-homogeneous process. It has already been shown that collapse of even a single isolated elastic ring is not instantaneous~\cite{djondjorov2011analytic} - complete at $P_c$ - but goes progressively through ovalisation to collapse shape over the range $P_c$ to 1.5$P_c$, see also Fig.3 in~\cite{2017-Torres-Dias}.\\

\section{Conclusion}
We propose a simple theoretical model to determine the stability domain of carbon nanotubes as a function of their diameters and their number of walls. The geometrical stability limits at ambient, as well as collapse pressures for arbitrary number of walls are characterized from the long range Van der Waals interactions, introduced into the modified L\'evy-Carrier equation, formulated for tubes at the nanoscale. The model proposed in this work is validated by numerical simulations at the nanoscale, as well as experiments. We have thus found that depending on the number of tube walls, nanotubes show a maximum collapse pressure ranging from $\sim$13 to 18 GPa with an inner-tube diameter of $d_0$=0.56nm. It is noteworthy that our model fits experimental data despite neglecting multi-wall nanotubes inter-layer friction. When $d$ is smaller than 0.56nm, the collapse pressure drops right down, due to the strong tube curvature, resulting in unstable nanotubes. For large tube diameters, the collapse pressure decreases, corresponding to the Van der Walls interactions that tend to favor the collapse. From the collapse phase diagrams plotted with the model presented in this work, we have shown that the L\'evy-Carrier equation (originally established for macroscopic tubes) is compatible with tube diameters of a few nanometers, and depending on the number of tube walls. This is an important result underlying that the collapse process of most of the common nanotubes produced by standard experimental techniques takes place as in macroscopic tubes, linking behavior at the nano- to the macroscale. This behavior was verified comparing numerical simulations with experiments at the nanoscale and at the macroscale, where nanotubes were replaced by polymer O-rings. We think that such an analogy maybe interesting in order to study mechanical deformation of nanotubes under pressure more easily, or more complex porous systems as Metal Organic Frameworks (MOF), zeolites, or even disordered porous materials as kerogen.

\section{Methods}
\subsection{Numerical simulations}
Density functional tight-binding calculations were performed using the DFTB software package~\cite{aradi2007} with the matsci-0-3 parameter~\cite{frenzel2005}. This algorithm was used only for small tube diameters ranging from $d_0$=0.5 to 1.4nm, due to its expensive computational time. In this approach, the Kohn-Sham density-functional theory is approximated with fitted integrals from reference calculations. The method increases simulations efficiency compared to density functional theory (DFT), while keeping "a priori" a better accuracy compared to the empirical approaches. The C-C Slater-Koster parameters implemented in this work have been extensively used for CNT simulations and can be found elsewhere~\cite{frenzel2005}. In  Fig.\ref{fig1}.a, we have determined $P_c$ for dozen of armchair SWCNT. For each pressure, a random displacement of 0.002nm is applied on each atom, and both atomic positions and cell vectors were optimized until the magnitude of all forces became smaller than 10$^{-4}$ Ha/Bohr. In this process, $P$ was increased by steps of 0.2~GPa, up to the tube collapse. This phenomenon is generally found to be abrupt, and can be easily identified from a discontinuity in the enthalpy as a function of $P$, corresponding to the transformation in a collapse shape. In some rare cases, especially for small $d_0$, the discontinuity is not visible, and $P_c$ was determined by eye, i.e., we assigned $P_c$ to the first collapsed geometry found.\\

\begin{figure*}[htp]
	\centering
	\includegraphics[width=.99\linewidth]{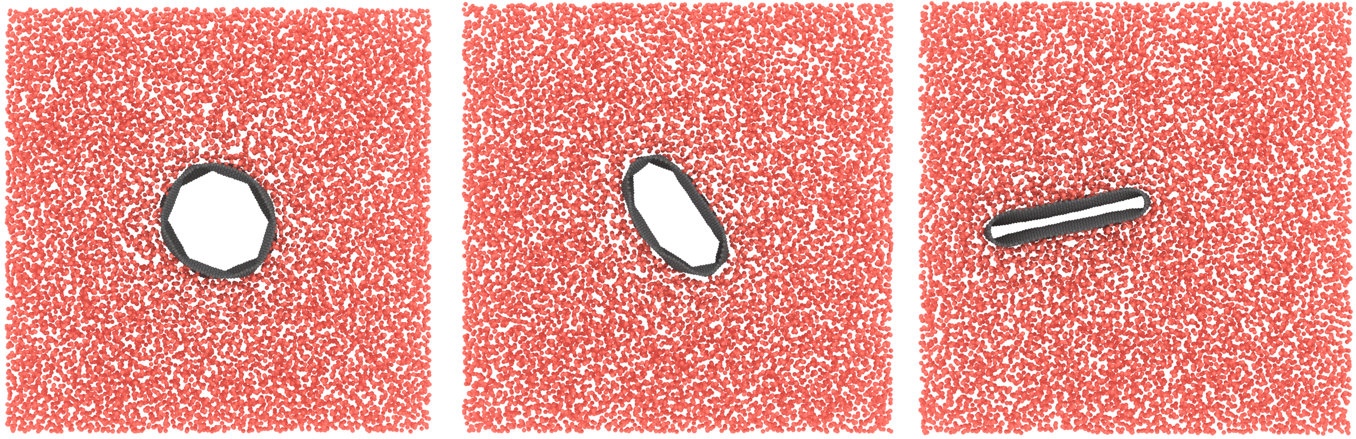}
	\caption{Shape evolution of a SWCNT with $d_0\sim$5~nm immersed into an argon bath at a pressure of P$\sim$0.5~MPa. The two first snapshots correspond to out of equilibrium configurations. The last configuration corresponds to an equilibrium state in a collapse shape.}
	\label{fig6}
\end{figure*}

\noindent A second set of simulations using an MD algorithm were performed to study larger tube diameters. MD simulations were conducted for systems with SWCNT immersed in an argon bath, in order to transmit pressure to the nanotubes. Inter-atomic interactions (Ar-Ar and Ar-C) were modelled by a (12-6) Lennard-Jones potential (LJ), with a cutoff fixed at 2~nm,
\begin{equation}
U = 4\ \epsilon \left[ \left(\frac{\sigma}{r}\right)^{12} - \left(\frac{\sigma}{r}\right)^{6}\right],
\end{equation}
where $\sigma$ corresponds to the atomic diameter, $r_{ij}$ is the inter-atomic distance, and $\epsilon_{ij}$ is the interaction energy between two atoms $i$ and $j$. The LJ parameters (given in Table \ref{tbl:LJ}) for the interactions of the different species are determined with the Lorentz-Berthelot mixing rule as, 
\begin{equation}
\sigma_{ij}=\left(\frac{\sigma_i\ +\ \sigma_j}{2}\right) \mbox{\hspace{1.0cm} and \hspace{1.0cm}} \epsilon_{ij}=\sqrt{\epsilon_i \epsilon_j},\\
\end{equation}
\begin{table}
  \caption{Lennard-Jones parameters for argon and carbon atoms.}
  \label{tbl:LJ}
  \begin{center}
  \begin{tabular}{lll}
	\hline
    Atoms & \hspace{1cm} $\sigma$ (nm) & \hspace{1cm} $\epsilon$ (meV)\\
    \hline
    \hline
    Ar-Ar    & \hspace{1cm} 0.341 & \hspace{1cm} 10.3\\
    \hline
    C-C    & \hspace{1cm} 0.336  & \hspace{1cm} 2.4\\
    \hline
    Ar-C    & \hspace{1cm} 3.382  & \hspace{1cm} 5\\
	\hline
  \end{tabular}
  \end{center}
\end{table}
\noindent The C-C interactions were modelled with the long range bond order potential AIREBO~\cite{stuart}. This potential models both the carbon covalent bonds, and long range vdW interactions with a cutoff fixed at 2~nm.\\
The first step of simulations consists in adsorbing Ar molecules in a bulk phase with the grand canonical Monte Carlo algorithm GCMC ($\mu_{Ar}$,$V$,$T$), where $\mu_{Ar}$ is the chemical potential of Ar, $V$ the volume of the simulation box at $T$=300K. GCMC was thus performed to generate a pure argon bulk configuration at pressures of about P=2~GPa. A tube (or a bundle) is then inserted inside the bulk configuration, and the inner part of the tube is cleared of argon atoms. MD simulations are then performed in the isothermal-isobaric ensemble ($N$,$P$,$T$), and P$_c$ is determined from enthalpy as a function of $P$. When tube collapse occurs, the tube energy changes abruptly, and the collapse pressure can be easily identify. In Fig.\ref{fig6}, we show the tube shape evolution, for a SWCNT with $d_0\sim$5~nm at P$\sim$0.5~MPa. As can be seen, the tube goes from a circular to a collapse shape, corresponding to its equilibrium state in such thermodynamic conditions. For all simulations performed in this work, we used zigzag SWCNT.

\subsection{Experiments}
\begin{figure*}[htp]
    \begin{center}
	\includegraphics[width=0.5\linewidth]{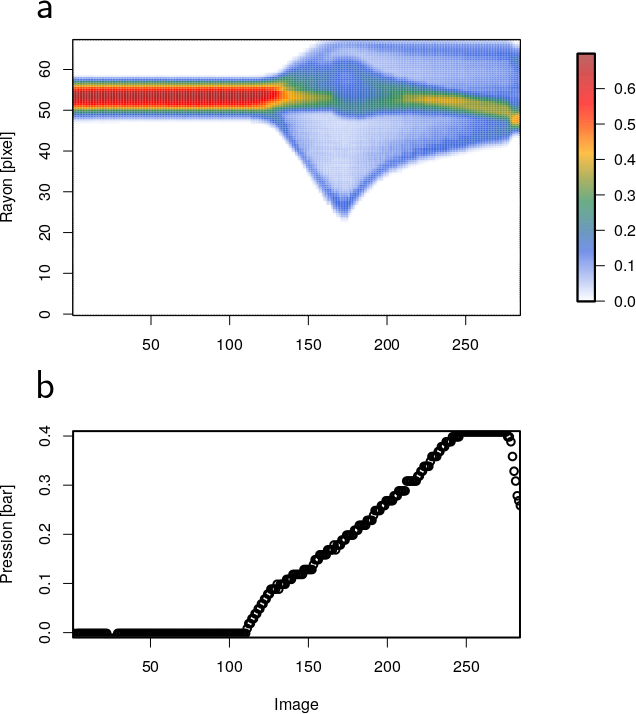}
	\caption{Determination of the collapse pressure for individual O-rings from video image analysis during the compression following the scheme of Fig.\ref{fig4}.a. \textbf{a.} Time  evolution of O-ring radial sectors, measured in pixels from the image. The color scale represents the radial distribution of the pixels detected. The higher this value, the more the O-ring presents a circular shape. On the contrary, the lower this value, the more the O-ring deviates from a circular shape, corresponding to a collapse situation. \textbf{b.} Pressure measured from the manometer in image analysis. The correlation with (a) allows to determine the collapse pressure. See videos in the supplementary material.}
	\label{fig7}
    \end{center}
\end{figure*}

In order to mimic the elastic properties of the radial buckling of SWCNT, we have used toroidal elastomer gaskets (O-ring), usually used for applications as vacuum seals. O-rings were placed between two transparent plates of PMMA (poly(methyl methacrylate)) with 25mm thickness, while a larger gasket diameter was used to create a cavity around the smaller one. We then used dynamometric keys in order to ensure a uniform and weak pressure over the O-ring. Both O-rings and internal plate surfaces were oiled in order to reduce the friction. Holes were then drilled into the top plate and were used to connect a bicycle pump and a manometer to vary and monitor the pressure respectively, Fig.\ref{fig7}. A schema and a picture of the experiment is shown in Fig.\ref{fig4}.a. Doing so, we was able to observe the effect of a radially applied pressure on O-rings for different diameters ($d_0$=11.25, 13.7, 16.25 and 28.75~mm), and taking care that displacements are constrained into the plane. In order to validate that O-rings can be compared to \textit{simili} of a SWCNT in the consistent LC domain, we have first shown that they verify the macroscopic LC approach itself, i.e., that the collapse pressure is inversely proportional to the cube of the torus diameter, Fig.\ref{fig4}.b.\\

\begin{figure*}[htp]
	\centering
	\includegraphics[width=1.\linewidth]{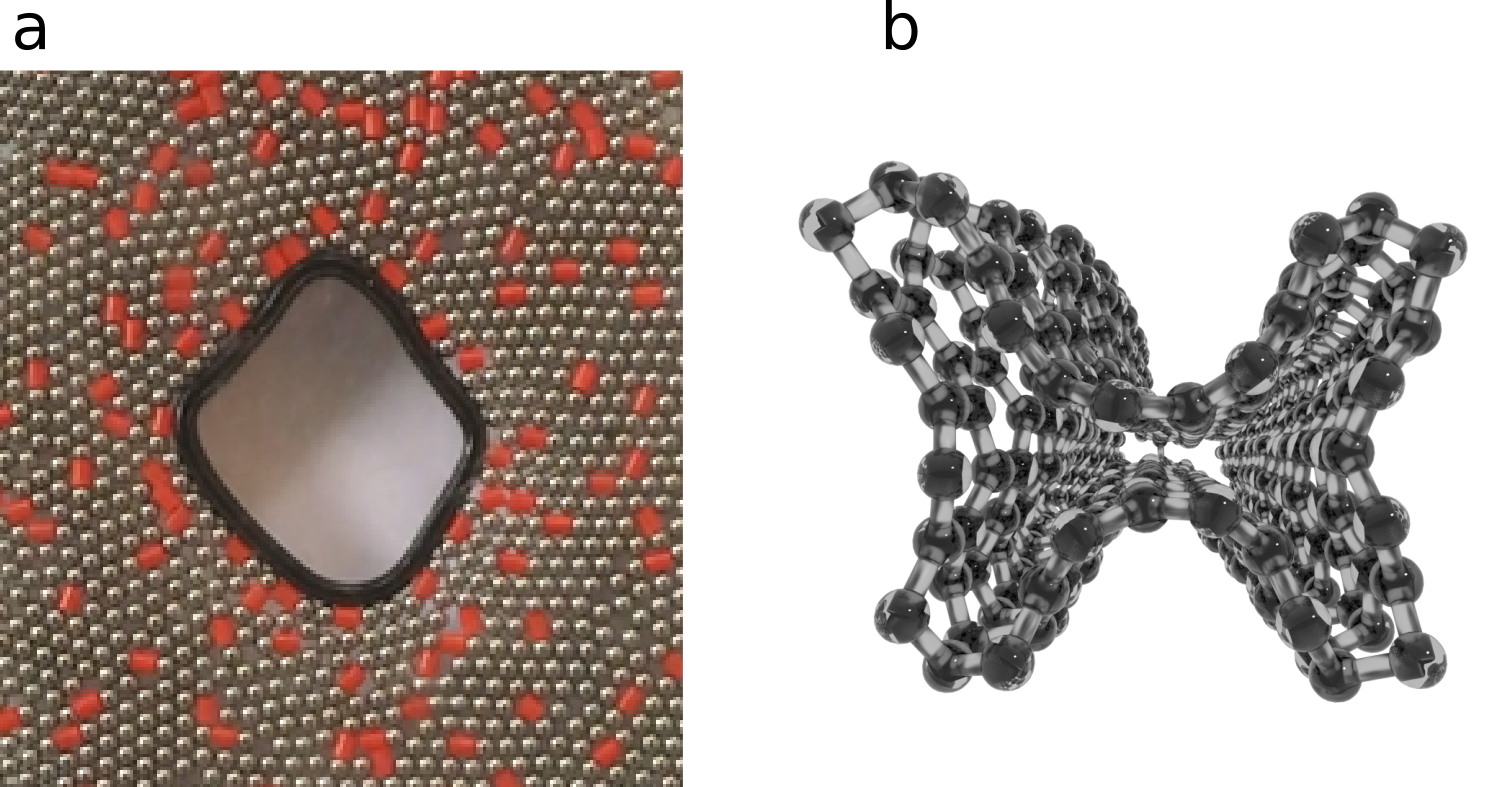}
	\caption{\textbf{a.} Four-lobe collapse of an O-ring obtained using a mixed medium of ball bearings and plastic beads (red colour). The mixed medium leads to the formation of smaller crystalline domains in the pressurized transmitting medium. \textbf{b.} Numerical simulation of tube collapse with a metastable state corresponding to a four-lobe shape (the pressure transmitting medium is not shown in this snapshot).}
	\label{fig8}
\end{figure*}

\noindent The second experiment conducted in this work is devoted to O-rings deformation using a pressure-transmitting medium that mimics the situation where SWCNT are immersed in a high-pressure ($>$1GPa) argon bath. To do this, we have used bearing balls enclosed by a guitar string, used to increase the pressure on the O-ring surfaces. As shown in the Fig.\ref{fig8}.a, when the string loop is tightened, the balls are well organised in a close-packed hexagonal structure. This crystalline medium shows grains as in a polycrystalline solid. Note that the grain structure induces rugosity at the surface of O-rings that replicates the situation at the nanoscale, see Fig.3 in~\cite{2017-Torres-Dias}. In order to limit such spurious effects, we have created a mixed medium by disseminating other types of particles as plastic beads into the ball bearing media, Fig.\ref{fig8}.a. Despite this trick, we was not able to collapse this O-ring due to the remaining grain domains, preventing from collapse. We show what should be such a four-lobe geometry in a fully collapse shape by MD simulations in Fig.\ref{fig8}.b.

\section*{Acknowledgments}
We acknowledge the platform PLECE of the University de Lyon and iLMTech (CNRS and University Claude Bernard Lyon 1). Y. Magnin gratefully acknowledges the Computational Center of Cergy-Pontoise University (UCP) for the computational time. A. San-Miguel and D. J. Dunstan acknowledge the support of the 2D-PRESTO ANR-19-CE00-0027 project.
\clearpage
\bibliography{Collapse-O-rings}

\end{document}